\documentstyle[12pt]{article}
\input psfig

\newcommand{\be}{\begin{equation}}
\newcommand{\ee}{\end{equation}}
\newcommand{\bea}{\begin{eqnarray}}
\newcommand{\eea}{\end{eqnarray}}
\newcommand{\half}{{\scriptstyle{{1\over 2}}}}

\newcommand{\trhf}{{\scriptstyle{{3\over 2}}}}

\newcommand{\Tr}{\mbox{\,Tr\,}}
\newcommand{\cF}{{\cal F}}
\newcommand{\cM}{{\cal M}}
\newcommand{\cX}{{\cal X}}
\newcommand{\cN}{{\cal N}}
\newcommand{\cK}{{\cal K}}

\newcommand{\cO}{{\cal O}}
\newcommand{\cV}{{\cal V}}
\newcommand{\pa}{{\partial}}
\newcommand{\RE}{{\Re e}}

\def\Journal#1#2#3#4{{#1} {#2} (#4) #3}
\def\NPB{{Nucl. Phys.} B}
\def\PLB{{Phys. Lett.} B}

\def\CMP{Comm. Math. Phys.}

\def\PRD{{Phys. Rev.} D}

\newcommand{\basispl}{
   \put(-.5,-.5){\line(1,0){1}}
   \put(.5,-.5){\line(0,1){1}}
   \put(.5,.5){\line(-1,0){1}}
   \put(-.5,.5){\line(0,-1){1}}}
\newcommand{\plaq}{\setlength{\unitlength}{.5cm}\raisebox{-.2cm}{
   \begin{picture}(1.2,1.2)(-.6,-.6)
   \basispl
   \put(-.5,-.5){\circle*{.2}}
   \put(-.5,.5){\circle*{.2}}
   \put(.5,-.5){\circle*{.2}}
   \put(.5,.5){\circle*{.2}}
   \put(.5,0){\vector(0,1){0}}
   \put(-.6,-.6){\makebox(0,0)[tr]{\footnotesize $x$}}
   \put(-.55,0){\makebox(0,0)[r]{\footnotesize $\nu$}}
   \put(0,-.60){\makebox(0,0)[t]{\footnotesize $\mu$}}
   \end{picture}}}
\newcommand{\twooneplaq}{\setlength{\unitlength}{.5cm}
   \raisebox{-.2cm}{
   \begin{picture}(2.2,1.2)(-1.1,-.6)
   \put(-1,-.5){\line(1,0){2}}
   \put(-1,.5){\line(1,0){2}}
   \put(-1,-.5){\line(0,1){1}}
   \put(1,-.5){\line(0,1){1}}
   \multiput(-1,-.5)(1,0){3}{\circle*{.2}}
   \multiput(-1,.5)(1,0){3}{\circle*{.2}}
   \put(-1.1,-.6){\makebox(0,0)[tr]{\footnotesize $x$}}
   \put(-1.05,0){\makebox(0,0)[r]{\footnotesize $\nu$}}
   \put(-.45,-.60){\makebox(0,0)[t]{\footnotesize $\mu$}}
   \put(1,0){\vector(0,1){0}}
   \end{picture}}}
\newcommand{\twoplaq}{\setlength{\unitlength}{1cm}\raisebox{-.5cm}{
   \begin{picture}(1.2,1.2)(-.6,-.6)
   \basispl
   \put(-.5,-.5){\circle*{.1}}
   \put(-.5,.5){\circle*{.1}}
   \put(.5,-.5){\circle*{.1}}
   \put(.5,.5){\circle*{.1}}
   \put(0,-.5){\circle*{.1}}
   \put(0,.5){\circle*{.1}}
   \put(.5,0){\circle*{.1}}
   \put(-.5,0){\circle*{.1}}
   \put(.5,-.2){\vector(0,1){0}}
   \put(-.55,-.55){\makebox(0,0)[tr]{\footnotesize $x$}}
   \put(-.55,-.2){\makebox(0,0)[r]{\footnotesize $\nu$}}
   \put(-.2,-.55){\makebox(0,0)[t]{\footnotesize $\mu$}}
   \end{picture}}}
\newcommand{\linkhmu}{\setlength{\unitlength}{.5cm}\raisebox{-.2cm}{
   \begin{picture}(1.2,1.2)(-.6,-.6)
   \put(.5,0){\line(-1,0){1}}
   \put(0,0){\vector(1,0){0.1}}
   \put(-.5,0){\circle*{.2}}
   \put(-.35,-.25){\makebox(0,0)[tr]{\footnotesize $x$}}
   \put(0.4,-.3){\makebox(0,0)[r]{\footnotesize $\mu$}}
   \end{picture}}}

\begin{document}
\hfill HD-THEP-96-45
\vskip-1mm
\hfill INLO-PUB-21/96
\vskip4mm
\begin{center}
{\LARGE{\bf{\underline{One-loop anisotropy for improved actions}}}}\\
\vspace{1cm}
{\large Margarita Garc\'{\i}a P\'erez${}^a$ and Pierre van Baal${}^b$
} \\
\vspace{1cm}
${}^a$ Institut f. Theoretische Physik, University of Heidelberg,\\
D-69120 Heidelberg, Germany.\\ 
%\vspace{5mm}
${}^b$ Instituut-Lorentz for Theoretical Physics, University of Leiden,\\
PO Box 9506, NL-2300 RA Leiden, The Netherlands.\\ 
\end{center}
\vspace*{5mm}{\narrower\narrower{\noindent
\underline{Abstract:} 
We determine the one-loop correction to the anisotropy factor for the square
Symanzik improved lattice action, extracted from the finite volume effective 
action for SU($\cN$) gauge theories in the background of a zero-momentum gauge 
field. The result is smaller by approximately a factor 3 than the one-loop
correction for the anisotropic Wilson action. We also comment on the 
Hamiltonian limit.
}\par}
\section{Introduction}

Improved actions~\cite{sym,luwe} have become a frequently used tool for doing 
Monte Carlo simulations. Recently the square Symanzik improved action was 
introduced~\cite{minn}, motivated by the desire to simplify perturbative 
calculations. From the numerical point of view this new improved action is 
not expected to be more optimal in removing lattice spacing errors, although 
simulations~\cite{us,taro} suggest it is not performing much worse than 
the L\"uscher-Weisz choice~\cite{luwe} of the improved action.

The square Symanzik action was introduced to allow for a simple background 
covariant gauge condition. The background field calculation is particularly 
suited for computing the renormalized coupling constant, not only in the 
continuum~\cite{tho}, but also on the lattice with, or without 
anisotropy~\cite{hada,kars}. Improved anisotropic lattices are used both 
for thermodynamics~\cite{taro} and for extracting glueball masses on 
very coarse lattices~\cite{morn}. In both cases the aim is to enhance the 
resolution in the time direction. Also for the square Symanzik action
anisotropy was introduced and used in Monte Carlo simulations~\cite{taro}.
This has motivated us to compute the one-loop correction to the anisotropy 
factor for this improved action, as it requires only a minor modification 
in the calculation already performed to compute the Lambda parameter
for its isotropic version.

We will perform the background field calculation for a finite volume at 
arbitrary anisotropy $\xi$, using the methods followed for the isotropic 
Wilson action~\cite{vbre}. Our results will include the Wilson action and 
the square Symanzik improved action. The one-loop correction to the 
anisotropy factor for the Wilson action was computed before by 
Karsch~\cite{kars} in an infinite volume. Our finite volume calculation 
nicely confirms the universality of these results. 

\section{The anisotropic square Symanzik action}
For the Wilson action the anisotropy was introduced as follows
($\linkhmu\equiv U_\mu(x)\in{\rm SU}(\cN)$):
\be
S_W(\xi)=\frac{1}{g_0^2}\sum_x\left[\eta\xi^{-1}\sum_{i>j>0}P_{ij}
+\eta^{-1}\xi\sum_{i\neq0}P_{0i}\right]\,,\ P_{\mu\nu}=2\RE\Tr(1-\plaq\ ),
\ee
where ($\beta\equiv 2\cN/g_0^2$ for SU($\cN$))
\be
\eta(\xi,g_0)=1+\eta_1(\xi)/\beta+\cdots
\ee
is required to guarantee that the symmetry of interchanging space and time 
is restored (in the infinite volume and continuum limit hence restoring 
Lorentz, or rather O(4), invariance). As we have changed the discretization 
of the theory, also the Lambda parameter belonging to the running coupling 
will depend on the anisotropy parameter. Both $\eta_1(\xi)$ and $\Lambda(\xi)$ 
can be determined from a one-loop calculation. In this paper we will use the 
notation $\xi(g_0)\equiv\eta^{-1}(\xi,g_0)\xi$, sometimes in the 
literature also denoted by $\gamma$. In the following $\xi(g_0)$ will be 
denoted by $\xi$ for short; from the context it should be clear when $\xi$ 
indicates the tree-level value. 

As was formulated in ref.~\cite{taro}, one can similarly introduce anisotropy 
for a tree-level improved action, 
\be
S(\{c_i\})\equiv\sum_x\RE\Tr\sum_{\mu\neq\nu}\frac{\xi_{\mu\nu}}{g_0^2}\Biggl\{
c_0\left(1-\plaq\ \right)+2c_1\left(1-\twooneplaq\ \right)+c_4\left(1-\twoplaq
\ \right)\Biggr\},
\ee
with 
\be
\xi_{\mu\nu}=\xi_\mu\xi_\nu,\quad \xi_i=\xi^{-\half},\quad \xi_0=\xi^{\trhf}.
\ee
We wish to emphasize that the issue here is not to improve this action 
beyond tree-level. It would involve the extra non-planar Wilson loops that 
also appear in the isotropic case~\cite{luwe}. Its coefficients, as well as 
the one-loop corrections to $c_0$, $c_1$ and $c_4$ will be doubled in number 
due to the anisotropy. After eliminating redundancies extra parameters 
will have to be determined, one of which can be related to $\eta_1$. However,
additional physical quantities are required to fix all one-loop coefficients, 
making this a less than straightforward generalization from the isotropic 
case~\cite{luwe,wei,snip}. The renormalization of the anisotropy parameter is, 
however, determined by requiring the restoration of the space-time symmetries 
in the continuum limit, and can therefore be addressed without computing the 
one-loop corrections to the improvement coefficients. 

We impose the renormalization condition not directly by the requirement to 
restore the space-time symmetries, but rather by comparing the finite volume 
effective action in a zero-momentum background field derived from the 
anisotropic lattice action in eq.~(3) with the result for the isotropic 
lattice action. We may also compare with the result obtained from dimensional 
regularization in the continuum. We will study the one-parameter family of
actions defined
by
\be
c_0=1/(1+4z)^2,\ c_1=z c_0,\ c_4=c_1^2/c_0,
\ee
where $z=0$ corresponds to the Wilson action and $z=-1/16$ corresponds
to the square Symanzik action, which is improved at tree-level to
second order in the lattice spacing.

The relation $c_4=c_1^2/c_0$ allows for a simple background field covariant 
gauge condition that is easily generalized to the anisotropic case.
\be
\hat\cF_{gf}\equiv\sqrt{c_0}\sum_{\mu}\xi_\mu\hat D_\mu^\dagger\left(1+z(2
+\hat D_\mu^\dagger)(2+\hat D_\mu)\right)\hat q_\mu(x)=0.
\ee
The background covariant derivative is given by $\hat D_\mu\Phi(x)\equiv 
\hat U_\mu(x)\Phi(x+\hat\mu)\hat U^\dagger_\mu(x)-\Phi(x)$ and the 
quantum fluctuations are parametrized as $U_\mu(x)=e^{\hat q_\mu(x)}
\hat U_\mu(x)$ for a lattice background field $\hat U_\mu(x)$. The free
propagators in this gauge (at $\hat U_\mu(x)\equiv 1$) are given by
\bea
{\rm Ghost}:&&P(k)=\frac{1}{\sqrt{c_0}\sum_\lambda\xi_\lambda\left(4\sin^2(
k_\lambda/2)+4z\sin^2k_\lambda\right)}\ ,\nonumber\\
{\rm Vector}:&&P_{\mu\nu}(k)=\frac{P(k)\delta_{\mu\nu}}{\sqrt{c_0}\,\xi_\mu\!
\left(1+4z\cos^2(k_\mu/2)\right)}\ .
\eea

\section{Background field calculation} 
We compute on a lattice of size $N^3\times\infty$ the effective action for a 
dynamical (i.e. time dependent) zero-momentum non-Abelian background field, 
$\hat U_j(x)\!\equiv\!\exp(\hat c_j(t))\!\equiv\!\exp(c^a_j(t)T_a/N)$ and 
$\hat U_0(x)\equiv 1$ (the anti-hermitian generators $T_a$ are normalized as 
$\Tr(T_aT_b)=-\half\delta_{ab}$). It is obtained by intergrating out all 
non-zero momentum modes. No integration over the zero-momentum quantum modes 
is included, which for a dynamical background field would lead to breakdown 
of the adiabatic approximation near $c=0$, where the classical potential is 
quartic~\cite{lush}. We will follow closely the methods developed for the 
isotropic Wilson action, described at great length before~\cite{vbre}. 
The effective action is given by
\be
\sum_ta_t\left\{\left(\frac{1}{g^2}\!+\!\alpha_1\!-\!\frac{\eta_1}{2\cN}
\right)\frac{\left(c^a_i(t\!+\!1)\!-\!c_i^a(t)\right)^2}{2a_t^2L^{-1}}\!
+\!\frac{1}{4L}\left(\frac{1}{g^2}\!+\!\alpha_2\!+\!\frac{\eta_1}{2\cN}
\right)\left(F_{ij}^a(t)\right)^2\!\!+\!V_1(c(t))\right\},
\ee
where $a_t=L/\xi N$ is the lattice spacing in the time direction, $L$ the 
physical size of the volume, $F_{ij}^a=\varepsilon_{abe}c_i^bc_j^e$ the
field strength and $V_1(c)$ is by definition the rest of the effective 
potential. All that is relevant to know is that at $\cO(c^4)$ its coefficients 
are fixed uniquely by an abelian background field, unambiguously separating 
$(F_{ij}^a)^2$ from $V_1(c)$. We have ignored terms that vanish in the 
continuum limit. Furthermore, the renormalization group to one-loop order 
implies $g_0^{-2}\!=\!-11\cN\log(a_s\Lambda)/24\pi^2$, where $a_s\!=\!L/N\!=\!
a_t\xi$ is the lattice spacing in the space directions. We have therefore 
introduced the renormalized coupling $g^{-2}\!\equiv\! g_0^{-2}\!-\!11\cN
\log(N)/24\pi^2\!=\!-11\cN\log(L\Lambda)/24\pi^2$.

We note that $\Lambda$, $\eta_1$, $\alpha_1$ and $\alpha_2$ depend on $\xi$ 
and $z$. Universality requires that physical quantities, as well as the 
background field effective action, are independent of these parameters in 
the continuum limit. This implies that both $(\alpha_1\!-\!11\cN\log(L\Lambda)/
24\pi^2\!-\!\eta_1/2\cN)$ and $(\alpha_2\!-\!11\cN\log(L\Lambda)/24\pi^2\!+\!
\eta_1/2\cN)$ are independent of $\xi$ and $z$. For isotropic lattices 
($\xi=1$), for which $\eta_1=0$, this implies that 
\be
\alpha_1-\alpha_1^c=\alpha_2-\alpha_2^c=11\cN\log(\Lambda/\Lambda^c)/24\pi^2
\quad(\xi\equiv1),
\ee
where $\alpha_i^c$ are the values for a fixed isotropic regularization, like 
dimensional regularization in the continuum, or the isotropic Wilson action.
For anisotropic lattices we follow Karsch~\cite{kars} by defining
\be
c_\tau(\xi)\equiv\alpha_1(1)-\alpha_1(\xi),\quad
c_\sigma(\xi)\equiv\alpha_2(1)-\alpha_2(\xi)
\ee
and one easily derives
\be
\eta_1(\xi)=\cN(c_\sigma(\xi)-c_\tau(\xi)),\quad \Lambda(\xi)/\Lambda(1)=
\exp\left(-12\pi^2[c_\sigma(\xi)+c_\tau(\xi)]/11\cN\right).
\ee
We can summarize these various relations between the Lambda parameters also as
\be
\Lambda(\xi,z)/\Lambda(\xi^\prime,z^\prime)=\exp\left(12\pi^2[\alpha_1(\xi,z)+
\alpha_2(\xi,z)-\alpha_1(\xi^\prime,z^\prime)-\alpha_2(\xi^\prime,z^\prime)]/
11\cN\right).
\ee

\section{Analytic results}
The coefficients $\alpha_1$ and $\alpha_2$ are determined by working out the 
determinants of the quadratic fluctuation operator. This section can be skipped
when one is interested in the numerical results only. To all orders in the 
background field and to quadratic order in the quantum field, using $c_0^a=0$
and $c^a_\mu(x+\hat\mu)=c^a_\mu(x)=c^a_\mu(t)$, we find ($k,\ell\in\{1,2\}$)
\bea
&&S_2=\frac{c_0}{4g_0^2}\sum_x\Tr\Biggl(\sum_{\mu,\nu,k,\ell}
\xi_{\mu\nu}z^{k+\ell-2}\Bigl\{
(\hat S^+_{k\hat\mu,\ell\hat\nu}(t)\!-\!2)\left(\hat D_{k\hat\mu}
\hat q_{\ell\hat\nu}(x)\!-\!\hat D_{\ell\hat\nu}\hat q_{k\hat\mu}(x)\right)^2-
\\&&\hskip3cm
\hat S^-_{k\hat\mu,\ell\hat\nu}(t)\bigl\{[q_{\ell\hat\nu}(x)\!+\!
\hat D_{\ell\hat\nu}\hat q_{k\hat\mu}(x),\hat q_{k\hat\mu}(x)\!+\!
\hat D_{k\hat\mu}\hat q_{\ell\hat\nu}(x)]+[\hat q_{\ell\hat\nu}(x),
\hat q_{k\hat\mu}(x)]+\nonumber\\&&\hskip4.6cm
(\ell\!+\!k\!-\!2)\hat D_{\ell\hat\nu}[\hat q_\mu(x),\hat D_\mu\hat q_\mu(x)]
\bigr\}\Bigr\}-4\Bigl\{\sum_{\mu,k}\xi_\mu z^{k-1}\hat D_{k\hat\mu}^\dagger
\hat q_{k\hat\mu}(x)\Bigr\}^2\Biggr).\nonumber
\eea
We have introduced the following convenient shorthand notations
\bea
&&\hat S^+_{k\hat\mu,\ell\hat\nu}(t)=2-\hat S_{k\hat\mu,\ell\hat\nu}(t)-
\hat S_{k\hat\mu,\ell\hat\nu}^\dagger(t),\quad\hat S^-_{k\hat\mu,\ell\hat\nu}
(t)=\hat S_{k\hat\mu,\ell\hat\nu}^\dagger(t)-\hat S_{k\hat\mu,\ell\hat\nu}(t),
\nonumber\\&&\hat S_{k\hat\mu,\ell\hat\nu}(t)=e^{k\hat c_\mu(t)}e^{\ell
\hat c_\nu(t+k\hat\mu)}e^{-k\hat c_\mu(t+\ell\hat\nu)}e^{-\ell\hat c_\nu(t)},
\eea
as well as ``doubled'' covariant derivatives and quantum fields ($q_{1\hat\mu}
(x)\!\equiv\! q_\mu(x)$ and $D_{1\hat\mu}\!\equiv\! D_\mu$)
\be
\hat D_{2\hat\mu}\Phi(x)=e^{2\hat c_\mu(t)}\Phi(x+2\hat\mu)
e^{-2\hat c_\mu(t)}-\Phi(x),\quad \hat q_{2\hat\mu}(x)=2\hat q_\mu(x)+
\hat D_\mu\hat q_\mu(x).
\ee
Therefore $\hat D_{2\hat\mu}=2\hat D_\mu+\hat D_\mu^2$ and the gauge fixing 
functional can be written as 
\be
\hat\cF_{\rm gf}(x)=\sqrt{c_0}\sum_{\mu,k}\xi_\mu z^{k-1}
\hat D_{k\hat\mu}^\dagger\hat q_{k\hat\mu}(x).
\ee
$S_2$ was obtained by adding $-\!\sum_x\!\!\Tr\hat\cF^2_{\rm gf}(x)/g_0^2$ to 
the action $S$. Under an infinitesimal gauge transformation, $\hat q_{k\hat\mu}
(x)$ transforms as $D_{k\hat\mu}\Phi(x)$ for $k=1,2$, giving for the ghost 
operator 
\be
\cM_{\rm gh}=\sqrt{c_0}\sum_{\mu,k}\xi_\mu z^{k-1}\hat D_{k\hat\mu}^\dagger
\hat D_{k\hat\mu}.
\ee

It is now straightforward to compute the functional determinants.  
Simplifications can be made due to the fact that we can split the one-loop 
correction in a purely kinetic part for which we can drop all terms of higher 
than second order in $c$ and a potential term for which the time dependence 
of the background field can be ignored. We find the following results
\be
\alpha_{s+1}(\cX)\equiv\alpha_d^{(s,2)}(\cX)+\xi^2\alpha_o^{(s,2)}(\cX)+
\xi^{2s}\alpha^{(2,s)}_o(\cX)+\alpha_{b}^{(s,2)}(\cX),
\ee
with $\cX=\{\xi,z,\cN,N\}$, $s=0,1$ and
\bea
\alpha_d^{(\mu,\nu)}(\cX)&=&\frac{11\cN}{24\pi^2}\log(N)+
\frac{\cN}{6\pi\xi^3 N^3}\sum_{\vec k\neq\vec 0}\int^\pi_{-\pi}dk_0 
\Bigl(d_\mu d_\nu-3q_\mu q_\nu\Bigr)P^2,
\\ \alpha_o^{(\mu,\nu)}(\cX)&=&
\frac{\cN}{8\pi\xi^3 N^3}\sum_{\vec k\neq\vec 0}\int^\pi_{-\pi}dk_0
\Biggl\{2(4-3p_\nu)(\frac{\zeta-d_\mu}{\cN^2}-\zeta)+
d_\mu(\zeta p_\nu^2+4(1-p_\nu))\Biggr\}P,
\nonumber\\ \alpha_b^{(s,\nu)}(\cX)&=&
\frac{(s\!+\!1)\cN}{12\pi\xi N^3}\sum_{\vec k\neq\vec 0}\int^\pi_{-\pi}\!\!dk_0~
\frac{\pa^2}{\pa k_\nu^2}\Biggl\{\frac{3\xi^{2s}}{4}\left[\zeta p_s^2+4(1-p_s)
\right]\log P+\xi^2\frac{\pa^2\log P}{\pa k_s^2}+d_s P\Biggr\}
\nonumber\\&&\hskip3.3cm +\frac{\pa}{\pa k_\nu}\left(6\zeta
P\frac{\pa \log p_\nu}{\pa k_\nu}\right),\nonumber
\eea
where $\zeta\equiv1+4z$, $P\equiv P(k,\xi,z)$ is the (rescaled) 
propagator, and $d_\mu$, $p_\mu$ and $q_\mu$ are simple trigonometric functions 
(momenta are given as $k\equiv(k_0,\vec k)=(k_0,2\pi\vec n/N)$, $n_i\in Z_N$)
\bea
&&\hskip-9mm P=\left[4\sin^2(\half k_0)\left(1\!+\!4z\cos^2(\half k_0)\right)
\!+\!\hat\omega^2\right]^{-1}\!\!\!,\ \hat\omega^2=\xi^{-2}\sum_i 4\sin^2(
\half k_i)\left(1\!+\!4z\cos^2(\half k_i)\right),\nonumber\\
&&\hskip-9mm d_\mu\!=\!\cos(k_\mu)\!+\!4z\cos(2k_\mu),\ p_\mu\!=\!\left[1\!
+\!2z(1\!+\!\cos(k_\mu))\right]^{\,-1}\!\!\!,\  q_\mu\!=\!p_\mu^{-1}(1\!+\!
\cos(k_\mu))(2\!-\!\zeta p_\mu)^2.
\eea

We note that in the continuum limit, $N\rightarrow\infty$, the total derivative
terms $\alpha_b$ will only get contributions from near 
$\vec k=\vec 0$. It can be shown that
\be
\lim_{N\rightarrow\infty}\alpha_b^{(s,2)}(\cX)=\cN\left(\frac{s+1}{
72\pi^2}+s\left(\frac{8a_4}{5}-\frac{1}{45\pi^2}\right)\right),\quad 
a_4=-(4\pi)^{-2}\cdot0.619331710\cdots
\ee
where $a_4$ is a constant introduced in ref.~\cite{lush}. In particular these 
boundary contributions are independent of $\xi$ and $z$ and drop out in the 
computation of the quantities in eqs.~(10-12). The remaining terms can be 
converted to {\em finite} integrals in the continuum limit, replacing 
$N^{-3}\sum_{\vec k}$ by $(2\pi)^{-3}\int_{-\pi}^\pi d^3\vec k$. 

For isotropic ($\xi=1$) actions, $\alpha_o^{(2,s)}=\alpha_o^{(s,2)}$ and 
$\alpha_d^{(s,2)}$ are independent of $s$. Consequently $\alpha_1\!-\!\alpha_2
\!=\!\cN(1/24\pi^2\!-\!8a_4)/5$ is independent of the regularization employed, 
cmp. eq.~(9). The non-vanishing value of this difference is a manifestation of 
the breakdown of Lorentz invariance in a finite physical volume.
At finite $N$ our results in eqs.~(18-19) are exact. All integrals over $k_0$ 
can be performed analytically, yielding sums over the $N^3-1$ non-zero spatial
momenta of {\em explicit} analytic expressions in $z$, $\xi$, $\cN$ and 
$\vec k$, which are readily evaluated numerically. It can be shown that as 
an expansion in $1/N$, terms linear {\em and for} $z=-1/16$ {\em (i.e. with 
improvement) quadratic} in the lattice spacing are absent. 
For $z=0$, where $p_\mu=\zeta=1$, $q_\mu=(1+d_\mu)$ and $d_\mu=\cos(k_\mu)$, 
dramatic simplifications occur. There are in particular for $z=0$ more
efficient ways to compute the coefficients, but we will not dwell on this
any further.

We list the following $k_0$ integrals required to evaluate the expressions in 
eq.~(19) ($y(\hat\omega$) is defined by $\hat\omega\equiv2\sinh(\half y)$ and 
$m>0$)
\bea
\delta_m^0&=&\frac{1}{2\pi}\int^\pi_{-\pi}dk_0~P^m=
\frac{1}{(m\!-\!1)!}\left(-\frac{\pa}{\pa \hat\omega^2}\right)^{m-1}
\left\{\frac{1}{2\sinh(y)\sqrt{1+4ze^{-y}}}\right\},\nonumber\\
\delta_m^1&=&\frac{1}{2\pi}\int^\pi_{-\pi} dk_0~\sin^2(k_0) P^m=
\frac{1}{(m\!-\!1)!}\left(-\frac{\pa}{\pa \hat\omega^2}\right)^{m-1}
\left\{\frac{e^{-y}}{1+4ze^{-y}+\sqrt{1+4ze^{-y}}}
\right\},\nonumber\\
\delta_m^{-1}&=&\frac{1}{2\pi}\int^\pi_{-\pi} dk_0~z^{-1}(1-p_0)P^m=
\frac{1}{(m\!-\!1)!}\left(-\frac{\pa}{\pa \hat\omega^2}\right)^{m-1}
\left\{4\hat\omega^{-2}\left(\frac{1}{\zeta+\sqrt{\zeta}}-\delta_1^1\right)
\right\},\nonumber\\
\delta_m^{-2}&=&\frac{1}{2\pi}\int^\pi_{-\pi} dk_0~p_0^2P^m=
\frac{1}{(m\!-\!1)!}\left(-\frac{\pa}{\pa \hat\omega^2}\right)^{m-1}\left\{
\frac{1+\zeta}{2\hat\omega^2\zeta^\trhf}+\frac{\zeta\delta_1^{-1}-4\delta_1^0
}{\hat\omega^2}\right\},\nonumber\\
\cK_m^1&=&\frac{1}{2\pi}\int^\pi_{-\pi}dk_0~d_0 P^m=
\frac{1}{(m\!-\!1)!}\left(-\frac{\pa}{\pa \hat\omega^2}\right)^{m-1}
\left\{(\half\hat\omega^2+\zeta)\delta_1^0-6z\delta_1^1-\half\right\},
\nonumber\\
\cK_m^2&=&\frac{1}{2\pi}\int^\pi_{-\pi}dk_0~q_0 P^m=
\frac{1}{(m\!-\!1)!}\left(-\frac{\pa}{\pa \hat\omega^2}\right)^{m-1}
\left\{\half \zeta^2\delta_1^{-1}-8z\delta_1^1\right\}.
\eea

\section{Numerical results}
By extrapolating in $1/N$ the explicit expressions for $\alpha_1$ and 
$\alpha_2$ on a finite lattice, we can extract their continuum limit to 
at least nine digit accuracy (we evaluate the momentum sums on lattices 
with $N=3$ to $N=99$). Our results are valid for arbitrary SU($\cN$), where
$\alpha_{1,2}$ can be written as $a\cN+b/\cN$. Using eqs.~(10-11) we 
reproduce results obtained by Karsch~\cite{kars} for the Wilson action. 
The ratio of the square Symanzik action Lambda parameter to the Wilson 
action Lambda parameter is obtained using either eq.~(9) or eq.~(12) with 
$(z,\xi)=(-1/16,1)$ and $(z^\prime,\xi^\prime)=(0,1)$. The result 
was already reported in ref.~\cite{minn}. One finds
\bea
\Lambda_{\rm sq}/\Lambda_{\rm W} &=& \left\{\begin{array}{ll}
4.0919901(1) & {\rm for\ }\cN=2,\\
5.2089503(1) & {\rm for\ }\cN=3, \end{array} \right.
\eea
agreeing with two alternative recent determinations based on the heavy-quark
potential and twisted finite volume spectroscopy~\cite{snip}.

In table 1 we give results for the square Symanzik action at some selected
values of $\xi$ between 1 and 20, likely to be of use in simulations, as well 
as for the Hamiltonian limit, $\xi=\infty$. The $\xi$ dependence is
illustrated in figure 1. Results for the Wilson action are given for 
comparison. We only present $\eta_1$ and $\Lambda(\xi)/\Lambda(1)$ for SU(2) 
and for SU(3), since one can use eq.~(11) to extract the values of 
$c_\tau$ and $c_\sigma$. These can furthermore be used to extract the 
results for any other number of colors, since 
\be
c_{\tau,\sigma}(\cN)=\frac{2(9-\cN^2)}{5\cN}c_{\tau,\sigma}(2)+
\frac{3(\cN^2-4)}{5\cN}c_{\tau,\sigma}(3).
\ee

We note that in all cases the value of $\eta_1$ is reduced by a factor of 
approximately 3 for the improved square Symanzik action as compared to the 
result for the Wilson action. Indeed for the simulations performed in 
ref.~\cite{taro} a reduction with approximately a factor 2.5 for the measured 
value of $\eta_1$ can be deduced (whereas the results obtained from the 
L\"uscher-Weisz and square Symanzik action agree within errors). We 
extracted $\eta_1$ using the one-loop truncation of eq.~(2). At the rather 
strong coupling employed in these simulations the measured values of $\eta_1$ 
themselves should of course not be expected to agree with the perturbative 
results~\cite{nucu}.

\section{Hamiltonian limit}
Here we briefly discuss an interesting feature of the Hamiltonian limit,
i.e. $\xi\rightarrow\infty$. It turns out that the potential $V_1(c)$ has
field dependent contributions that diverge in this limit. This divergence, 
however, only occurs for $z\neq0$, in particular for the improved square 
Symanzik action, i.e. at $z=-1/16$. At first sight this may seem puzzling. 
However one should notice that when modifying the action also  the measure 
of integration has to be corrected. Such a modification of the measure can 
of course be absorbed in the action, as is usually done and is of the same 
order as the one-loop corrections, giving rise to a term $\sum_{t,i}\delta 
V(c_i(t))$. Since the effective potential appears in the action as $\sum_t
a_t V_1(c(t))$, with $a_t=L/N\xi$, we conclude that the total 
contribution to the effective potential, due to correcting for the measure, 
is linear in $\xi$ (vanishing for $N\rightarrow\infty$) and given by 
$N\xi\sum_i\delta V(c_i)/L$. It is hence much more natural to redirect any 
terms linear in $\xi$ to the measure. 

To determine $\delta V$ we compute $V_1(c)$ by taking an abelian background 
field, suitably extended to the non-Abelian sector. For SU(2) this extension 
is achieved by substituting $C_i=\sqrt{\sum_a c_i^ac_i^a}$ in the Abelian 
background link variable $\hat U_j=\exp(\half iC_j\sigma_3/N)$. This can be 
generalized to arbitrary gauge groups following the methods described in 
ref.~\cite{carg}, but we will for the sake of presentation only consider 
the effective potential for SU(2). At $\cO(c^6)$ there are additional terms 
that vanish for Abelian background fields, but they do not concern us here 
(and have finite limits as $\xi\rightarrow\infty$). Along the lines described 
in ref.~\cite{minn} anisotropy is easily incorporated and one finds $V_1(c)=
V_1^{\rm ab}(\vec C)-V_1^{\rm ab}(\vec 0)$, where
\be
V^{\rm ab}_1(\vec C)=\frac{N\xi}{L}\sum_{\vec n\neq\vec 0\in Z_N^3}
\Biggl\{\sum_i\log\left(\lambda_i\right)+4{\rm asinh}\left(\frac{1}{\sqrt{
4|z|}}\,\sqrt{1+4z+\frac{\omega^2}{2\xi^2}+\frac{\omega}{\xi}
\sqrt{1+\frac{\omega^2}{4\xi^2}}}\ \right)\Biggr\},
\ee
with $\lambda_j(n_j,C_j)=1+4z\cos^2((\pi n_j+\half C_j)/N)$ and $\omega^2(
\vec n,\vec C)=\sum_j4\lambda_j\sin^2((\pi n_j+\half C_j)/N)$ (a more detailed 
derivation for the isotropic case will appear elsewhere~\cite{thes}). One can 
verify that this gives the correct continuum limit at fixed $\xi$. For the 
Wilson action ($z\!=\!0$) one finds $V^{\rm ab}_1(\vec C)=L^{-1}N\xi\sum_{
\vec n\neq\vec 0}4{\rm asinh}(\omega/2\xi)$ (the apparent divergence for 
$z=0$ can be shown to be field independent). At finite $N$ we find $\delta 
V(c_i)=\sum_{\vec n\neq\vec 0}\log\left(\lambda_i(n_i,C_i)/\lambda_i(n_i,0)
\right)$. The remainder, denoted by $\cV_1(c)$, can easily be shown to have 
a finite limit for $\xi\!\rightarrow\!\infty$. We conclude that the 
Haar-measure, $d\hat U_i(c)=N^{-1}(2\pi C_i)^{-2}\sin^2(C_i/2N)\prod_adc^a_i$, 
is to be corrected with a factor $\exp\left[-\delta V(c_i)\right]$, giving the 
{\em exact} measure at finite $N$ for deriving the effective Hamiltonian from 
improved actions. It is not too hard to show that up to exponential corrections
in $N$, we have $\exp[-\delta V(c_i)]=\left[1+4z\cos^2(C_i/2N)\right]/(1+4z)$.
Indeed at $z=-1/16$ the rescaled measure is flat to $\cO(c^3/N^3)$.

It is also interesting to point out that one finds $L \cV_1(c)=\gamma_1(
c_i^a)^2+\cO(c^4)$, with $\gamma_1=\gamma_1^c-2z/N\xi\sqrt{1+4z}+\cO(N^{-3})$. 
Since at $z\!=\!-1/16$ on-shell improvement should imply that spectral 
quantities have no $\cO(1/N^2)$ errors, the $1/N$ correction to $\gamma_1$ 
can be removed by a non-local field redefinition, as was explained to some 
detail in ref.~\cite{lat96}. The field redefinition is designed to remove the 
next-to-nearest couplings in the time direction (not listed in eq.~(8) since 
they are irrelevant in the continuum limit). This non-local effect disappears 
in the Hamiltonian limit, as it should.

\section{Conclusions}
We have calculated the one-loop correction to the anisotropy parameter
for the square Symanzik action, using a zero-momentum background field 
calculation in a finite periodic volume. For the Wilson case we retrieved 
the results in an infinite volume of Karsch~\cite{kars}, which is a rather 
non-trivial check of universality, since even in the continuum the periodic 
boundary conditions break the Lorentz invariance. We find that the size of 
the one-loop correction to the anisotropy is reduced, both for SU(2) and 
SU(3), by approximately a factor 3 when using the square Symanzik improved 
action instead of the Wilson action.

We have also discussed the ``Hamiltonian limit'' of the zero-momentum effective
theory, where the lattice spacing in the time direction is reduced to zero. 
We show how the integration measure is to be improved, defining the inner
product on the Hilbert space involved in extracting an effective Hamiltonian
from the effective action~\cite{kosu}.

\section*{Acknowledgements}

This work was supported in part by FOM and DFG. We are grateful to Ion
Stamatescu for persuading us to incorporate anisotropy in our one-loop
calculations and for useful discussions. PvB is grateful to Peter Lepage 
and Stephen Sharpe for hospitality at the workshop ``Improved actions for 
lattice QCD'' (INT, Seattle, 4-6 September, 1996) and for the stimulating
atmosphere created by all its participants.

\newpage
\def\mystrut{{\vrule height 11pt depth 4pt width 0pt}}
\def\ph{\hphantom{2}}
\hspace{4.5cm} Square  Symanzik\hspace{3.2cm}Wilson
\vskip.8mm
\hbox to \hsize{\hfil\vbox{\offinterlineskip
\halign{&\vrule#&\ $#\mystrut$\hfil\ \cr
\noalign{\hrule}
& &&\quad\xi&&\quad\quad\ \eta_1&&\ \Lambda(\xi)/\Lambda(1)&& &&\quad\quad
\eta_1&&\,\Lambda(\xi)/\Lambda(1)&\cr
height 4pt&\omit&&\omit&&\omit&&\omit&&\omit&&\omit&&\omit&\cr
\noalign{\hrule}
height 4pt&\omit&&\omit&&\omit&&\omit&&\omit&&\omit&&\omit&\cr
&     &&\ph1.25&&0.0268994230&&0.9448516628&& &&0.072639575&&0.941156329&\cr
&     &&\ph1.50&&0.0438407406&&0.9030990984&& &&0.120815052&&0.895220389&\cr
&     &&\ph1.75&&0.0553875971&&0.8715577166&& &&0.154865904&&0.862024175&\cr
&     &&\ph2.00&&0.0637309147&&0.8473816328&& &&0.180064348&&0.838519956&\cr
&     &&\ph2.25&&0.0700334250&&0.8285038242&& &&0.199377634&&0.821947171&\cr
&     &&\ph3.00&&0.0821772181&&0.7913271531&& &&0.236952649&&0.796282892&\cr
&     &&\ph4.00&&0.0909349958&&0.7647074013&& &&0.263881756&&0.786799000&\cr
&SU(2)&&\ph5.00&&0.0960771084&&0.7495355534&& &&0.279376900&&0.786585170&\cr
&     &&\ph6.00&&0.0994693690&&0.7398121769&& &&0.289388945&&0.789176074&\cr
&     &&\ph7.00&&0.1018786317&&0.7330678148&& &&0.296372263&&0.792513704&\cr
&     &&\ph8.00&&0.1036794982&&0.7281198326&& &&0.301513492&&0.795887850&\cr
&     &&\ph9.00&&0.1050771985&&0.7243357505&& &&0.305453471&&0.799053830&\cr
&     &&  10.00&&0.1061937831&&0.7213480410&& &&0.308567683&&0.801940305&\cr
&     &&  20.00&&0.1112064198&&0.7083043501&& &&0.322153959&&0.819059857&\cr
&   &&\ph\infty&&0.1162101357&&0.6957761241&& &&0.335019703&&0.843515849&\cr
\noalign{\hrule}
height 4pt&\omit&&\omit&&\omit&&\omit&&\omit&&\omit&&\omit&\cr
&     &&\ph1.25&&0.0761124472&&0.9441552990&& &&0.202232512&&0.940150646&\cr
&     &&\ph1.50&&0.1259027090&&0.9013716023&& &&0.339196380&&0.893219710&\cr
&     &&\ph1.75&&0.1609870225&&0.8690317960&& &&0.437758448&&0.859889187&\cr
&     &&\ph2.00&&0.1870579338&&0.8443511780&& &&0.511822337&&0.837010062&\cr
&     &&\ph2.25&&0.2072120386&&0.8252080564&& &&0.569337480&&0.821587813&\cr
&     &&\ph3.00&&0.2472872614&&0.7880636199&& &&0.683440912&&0.800832184&\cr
&     &&\ph4.00&&0.2772479446&&0.7621707435&& &&0.767394275&&0.798377544&\cr
&SU(3)&&\ph5.00&&0.2952330195&&0.7477916435&& &&0.816720193&&0.804338358&\cr
&     &&\ph6.00&&0.3072395475&&0.7387537514&& &&0.849056600&&0.812081537&\cr
&     &&\ph7.00&&0.3158270143&&0.7325745051&& &&0.871852492&&0.819692104&\cr
&     &&\ph8.00&&0.3222747701&&0.7280896992&& &&0.888772783&&0.826632621&\cr
&     &&\ph9.00&&0.3272941994&&0.7246877937&& &&0.901823656&&0.832804921&\cr
&     &&  10.00&&0.3313126485&&0.7220187467&& &&0.912193385&&0.838252399&\cr
&     &&  20.00&&0.3494266569&&0.7105270954&& &&0.958042934&&0.868813879&\cr
&   &&\ph\infty&&0.3675789970&&0.6996590739&& &&1.002502899&&0.910408485&\cr
\noalign{\hrule}}}\hfil}
\vskip5mm
{\narrower\narrower\noindent
Table 1: Results for $\eta_1$, the one-loop correction to the anisotropy $\xi$,
and the Lambda ratios for SU(2) and SU(3) Wilson and square
Symanzik improved lattice actions.\par}
\begin{figure}[htb]
\vspace{14cm}
\includegraphics{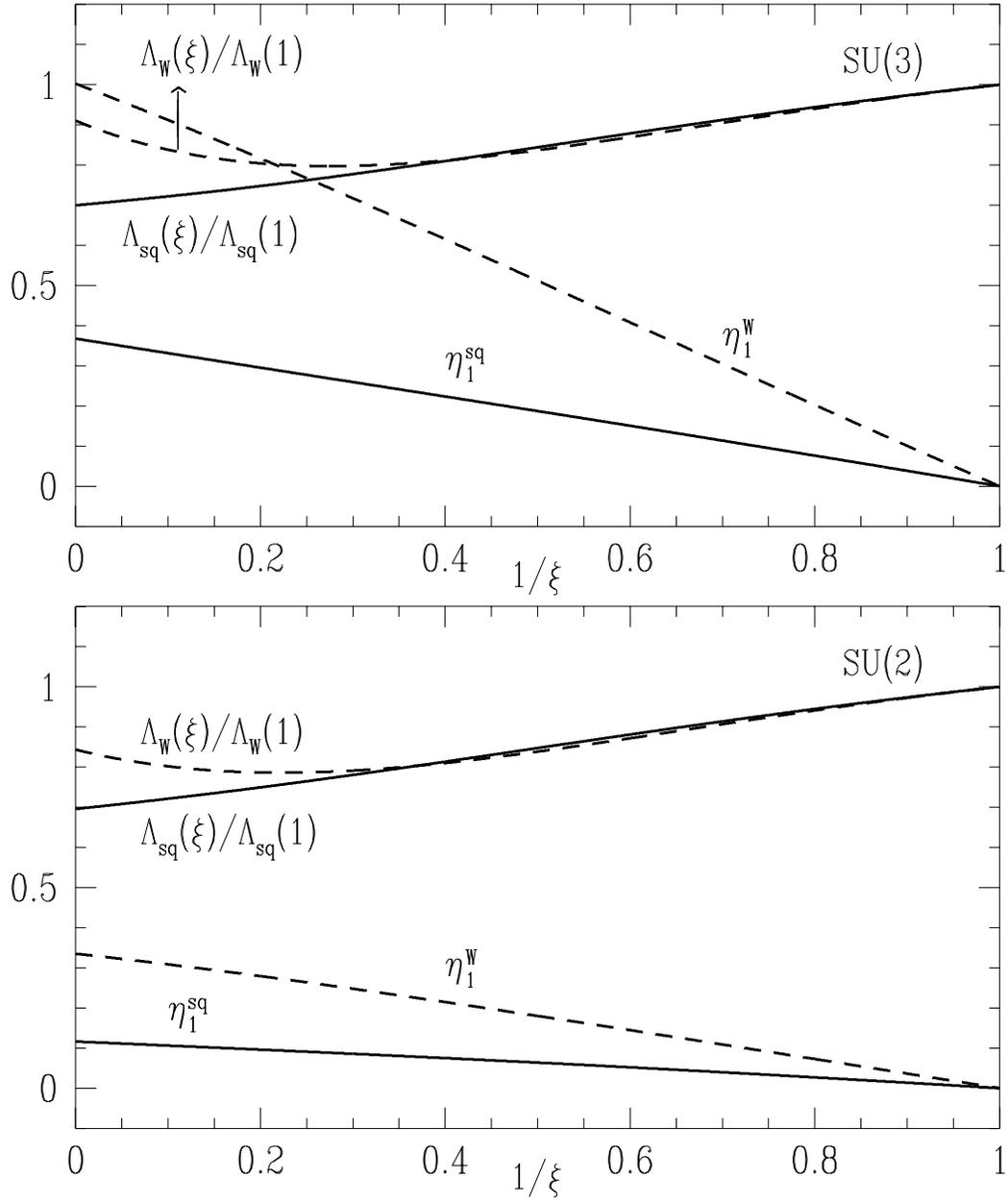}
\caption{Comparison between the square Symanzik (sq) and Wilson (W) action 
results for $\eta_1$ and $\Lambda(\xi)/\Lambda(1)$. Bottom figure for SU(2) 
and top for SU(3).}
\end{figure}
\end{document}